\documentclass[conference]{IEEEtran}
\usepackage{inconsolata}
\IEEEoverridecommandlockouts
\usepackage{cite}
\usepackage{amsmath,amssymb,amsfonts,amsthm}

\usepackage{algorithmic}
\usepackage{graphicx}
\usepackage{textcomp}
\usepackage{xcolor}
\usepackage{tabularray}

\usepackage{booktabs}

\usepackage[svgnames]{xcolor}
\usepackage[colorlinks,urlcolor=NavyBlue,citecolor=Maroon,linkcolor=DarkSlateGrey]{hyperref}
\usepackage{hyperref}

 \usepackage{todonotes}
\setuptodonotes{inline,backgroundcolor=brown!30,linecolor=black!30}

\usepackage{tikz}
\usetikzlibrary{automata,positioning,arrows.meta,backgrounds,fit,calc}
\tikzset{arr/.style={-Latex}}
\usepackage{pgfmath}
\usepackage{tikz-cd}

\usepackage{bm} 
\usepackage[a]{esvect} 

\usepackage{orcidlink}

\theoremstyle{plain}

 \theoremstyle{definition}

\newcommand{\CC}{\textsc{Con-Cat}}

\begin{document}

\title{What if we have 90 minutes only to teach programming?
\thanks{This project was funded in part by the Kakenhi grant 22K00015 by the Japan Society for the Promotion of Science (JSPS), titled `On progressing human understanding in the shadow of superhuman deep learning artificial intelligence entities' (Grant-in-Aid for Scientific Research type C, \url{https://kaken.nii.ac.jp/grant/KAKENHI-PROJECT-22K00015/}).}
}
\author{\IEEEauthorblockN{Attila Egri-Nagy}\IEEEauthorblockA{\textit{Human \& AI Center}\\\textit{Akita International University}\\ Akita, Japan \\ \orcidlinkf{0000-0001-7861-7076}}}
\maketitle


\begin{abstract}
  Programming is about automation in a wide variety of domains.
  Developing itself is one of those.
  As a side-effect, progress in automated coding may make people less willing to learn computer programming.
  This could become an issue, if the skill of computational problem solving is not only for the immediate economic benefit, but an important part of our knowledge about the world.
  We suggest that weakened incentives can be countered by lowering the entry barrier.
  We plan to shorten learning time by reducing the accidental complexity of the programming language and its runtime system.  
 We describe a session plan that introduces programming and computing fundamentals for novices, assuming only basic mathematical background.
  This requires a non-mainstream, functional and concatenative language.
  This language, \CC, is a by-product of research in category theory.
  It provides direct access to fundamental ideas like recursion and advanced ones like Gödel-encoding in an entertaining puzzle-like manner.
\end{abstract}

\maketitle



\begin{quote}
``\ldots  one of the most important aspects of any computing tool is its influence on the thinking habits of those that try to use it,''
  \end{quote}
\begin{flushright}The Humble Programmer
by
Edsger W. Dijkstra, 1972 Turing Award Lecture \cite{dijkstra1972humble}
\end{flushright}

\section{Introduction}
Automation is a driving force of computer programming.
Recently, the domain of automation has reached programming itself.
It is predicted that the activity of writing code will go extinct \cite{2023endofprogramming}.
Therefore, it is conceivable that in the future fewer people will learn how to develop computer programs.
In turn, this may lead to a reduction in the number of people understanding the technologies underpinning our everyday life.

The situation in programming education is similar to writing, which arguably the same skill as clear thinking \cite{zinsser1988writing}.
It is predicted, that even the number of people who can write will diminish, splitting the population into `writes and write-nots' \cite{paulgraham2024}.
Similarly, we may have an even more contrasted `codes and code-nots'.
This might turn out to be an overly pessimistic prediction. The utility of coding skills for the society is changing and it is actively debated.
We indeed hope that we are incorrect in forecasting a decline in engineering skills.
Still, it is better to be prepared.
Our cornerstone assumption is that losing knowledge is not a good thing.
We plan to counter the disappearance of economic incentives by making the barrier to advanced ideas lower.
This requires rethinking the basics, identifying the essentials, and making the activity attractive.

Going back to the absolute basics has been advocated before in the context of the service teaching of computer science \cite{selfie2017}.
That paper describes a complete self-referential runtime system for a \textsc{C}-like language.
It also invites the reader to think about what would be a bare minimum to teach other parts of computing.
Another successful approach is the ``From Nand to Tetris'' course \cite{2024nand2tetris}.
It involves hardware design as well.
A rather different take on teaching programming for all is Scratch \cite{2009scratch}.
It emphasizes narratives, creativity and social interaction, using programming as a means to achieve those goals.

This paper describes a blend of high-level functional and low-level machine code style of programming, aiming to stay close to the mathematical theories of computation.
Our focus here is on \emph{reducing learning time}.
Even if people decide to study programming, they may not be willing sustain the required effort for long.
To put it pessimistically, we might only have 90 minutes of student attention in the future, hence the title of this paper.
Can programming ideas be taught in a very short time?
Naturally, the master plan is to show enough to capture interest and ensure future attention.

\subsection{The origin story of \CC: a happy accident}

We were not set out to develop educational material.
The goal was to design a programming language for doing hierarchical decompositions in algebraic automata theory \cite{sgpoiddec}.
We needed an explicit semigroupoid representation, where concatenation corresponds to function composition.
The niche paradigm of concatenative languages can serve this role.
The design process and inspiring languages are described in \cite{egrinagy2025candarw}.
The closest ancestor is the \textsc{Joy}~programming language \cite{von2001joy}.

During the development it was intentional to do practical programming in the language, beyond examining its mathematical aspects.
Then, it was recognized that the unusual syntax  makes writing problems very puzzling. 
At the same time, the solutions are intellectually satisfying and very expressive.
These are the hallmarks of challenging course materials.

\subsection{Essential and Accidental Complexity}
The seminal software-engineering paper \cite{brooks1986no} distinguishes between two types of complexity arising in problem solving.
Roughly speaking, the \emph{essential complexity} belongs to the problem we want to solve, and there is no way to eliminate it.
On the other hand, the \emph{accidental complexity} comes from the tools we use.
Consequently, we can change it by switching to other tools.

For instance, using a computer requires knowing how to program and manage running applications.
The invention of high-level languages is a huge step in reducing this accidental complexity.
Of course, there always remains a bit of friction in using any tools.
Complete elimination is not possible, but we aim to minimize.

\subsection{Happiness as being in the zone}

It is a commonplace that we learn better when we enjoy the process.
In \cite{csikszentmihalyi1990flow}, this observation was made precise and confirmed by long-term psychological studies.
The \emph{optimal experience} is defined as \emph{``a sense that one's skills are adequate to cope with the challenges at hand, in a goal-directed, rule-bound action system that provides clear clues as to how well one is performing''}.
For a programming language, the complexity of the challenge can be fine-tuned by giving subsets of the language (e.g., \cite{felleisen2018design}).
We see this balancing at the edge of manageable complexity as the main driving force for learning.
Once a concept is familiar, it becomes boring, thus there is a need for proceeding to more challenging problems.
We design pathways in the course material to support this succession.

It is possible to have a culture where mathematical and problem solving is not directly tied to practical applications.
During the Edo period (1603–1867) in Japan, \emph{wasan} was developed a form of recreational but serious mathematics \cite{sakurai2018wasan}.

\subsection{The axiomatic method}

Ever since Euclid's Elements, the ideal mathematical knowledge representation is a formal system: \emph{axioms} and \emph{rules of inference}.
This is a form of compression: the true statements of the theory can be derived from the axioms.
Similar compression applies to programming languages: \emph{primitive programs} and \emph{composition}.
In concatenative languages, a library function definition is just a named sequence of programs. 
What are the language primitives and what library functions are built on top of those is an important design decision.
It affects execution speed, and more importantly for us, it determines how bootstrappable the language is. 

Since, we want to have a programming language that can be used even without a computer, just pen and paper computation, we want to have a core as small as possible.
Choosing the best set of primitive operators is not trivial.
The combinators are not independent.
They are often interconstructable \cite{2007concatenativeoperators}.

The minimality of the language definition may also serve a different purpose.
The \emph{permacomputing} movement \cite{2022permacomputing} advocates several ideas, including \emph{software longevity}.
If the runtime system is small, then it can be implemented on any existing or future systems, ensuring that code written today will work years later.

\subsection{The target level: Project Euler}

\textsc{Project Euler}\footnote{\url{https://projecteuler.net/}} is a website hosting a collection of mathematical problems that require both insight and computational power.
Typically the problems cannot be solved by using a brute force approach, and at the same time a good idea alone is not sufficient, as several cases need to be computed.
It is well recognized that the challenge problems are useful for both learning programming and mathematics \cite{2018projecteuler}.
Some problems from the project were used for testing \CC.
It is envisioned that a successful completion of the course material will make it possible to work on the challenge problems.

\section{The Course Material}

Keeping up with the promise of creating a very short course material, we present the key ideas of the tutorial here.
The reader can judge whether this material is sufficient enough for understanding the basic mechanisms of computation, or not.
Naturally, the time for reading the paper could be shorter than the planned 90 minutes, but it lacks all the interactions and communications of the live session.

Significant part of developing the material was about refining the notation for execution traces.
These are used for quizzes too.
Part of the stack or the code is replaced by a question mark.
The task is to find the missing part to make the execution consistent.

\subsection{Stack, Input Buffer and the Execution Model}

The \emph{stack} data structure appears in multiple ways in life: stacking coins, plates, or trays in a cafeteria.
The underlying mechanism is an intuitive idea: last thing in is the  first go out.
There is no need for a formal definition in an introduction.
The empty stack is denoted by a little box.
In the notation used below the stack's top is on the right its depth grows towards the left.
Here is a simple execution trace.
Time proceeds from top to bottom.
\begin{center}
  \includegraphics{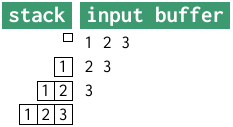}
\end{center}  
The \emph{input buffer} is on the right.
It is a sequence of \emph{tokens}.
A token is just a sequence symbols not including whitespace characters
The stack is on the left.
At one step, we pick the left-most token and evaluate it.
Basic data items just push themselves to the stack.
We can identify data literals with programs that push values to the stack.
This way everything is a program.

\emph{Operators} do something with the items on the stack.
For instance, \texttt{dup} pushes a new copy of, while \texttt{drop} discards the top value. 
\begin{center}
  \includegraphics{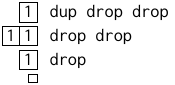}
\end{center}  
We can point out some common sense cases of errors.
We cannot duplicate or throw away the top of the stack if the stack is empty.

At this stage these stack manipulation operators seem to be of little use.
They become crucial in more complex computations.
However, experienced students are very quick to recognize that any stack state can be constructed easily: clearing the stack by dropping everything and then filling it up with the desired data items.
Eventually, there is a need to postulate that simply dropping and pushing is not a valid solution for a problem.

\subsection{Arithmetic}

Arithmetic is a solid base for learning about computation, since everyone is familiar with it. One plus one equals two.
\begin{center}
  \includegraphics{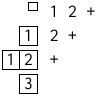}
\end{center}  
Since we use a stack, the arithmetic is postfix (Reverse Polish notation, RPN).
The operators follow the operands and they have fixed arity, therefore there is no need for parentheses.
This may be unusual for most people, but vintage calculators had this notation.
It has its advantages: faster and less error-prone operation \cite{RPNvsAN1980}.
In \CC~RPN is due to its mathematical origin and its simple parsing. 

Less experienced participants may have issues with `unseen actions'.
The above trace treats the addition atomic, while it has several steps.
It needs to pop the operands from the stack, carry out the additions in a separate \emph{workspace}, and push the result back to stack.

In the first quiz, we need to find the result of the computation.
The vertical dots indicate omitted steps to shorten the traces.
\begin{center}
  \includegraphics{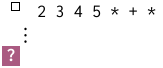}
\end{center}  
The problem simply requires playing out the evaluation steps.
The solution corresponds to the expression $(2*(3+(4*5)))$, which equals 46.

It is more interesting to turn the question around: finding not the result of the execution of some program, but constructing a program that yields the desired result from the given input.
\begin{center}
  \includegraphics{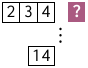}
\end{center}
How would such a program look like?
We have three data items, thus by using two arithmetic operators we can get down to a single number on the top of the stack.
Restricted to addition and multiplication, we have four possible such programs: \texttt{+ +} yields 9, while \texttt{+ *} computes $2*7=14$, the desired result.
By checking the possibilities systematically, we can find that \texttt{* +} is also a solution by $2+12=14$, and \texttt{* *} produces 24.
This is an excellent opportunity to reflect on the possibility that a problem may have several solutions.
Later, the diversity may be desirable for performance or readability reasons.

\subsection{Lists}

The fundamental collection data type of the language is the list.
It is inhomogeneous and we can join elements on both ends, unlike in \textsc{Lisp}.
The square brackets \texttt{[} and \texttt{]} are separate tokens.
When processing an opening bracket, we push an \emph{open list} to the stack.
It will swallow all the tokens until the closing bracket arrives.
\begin{center}
  \includegraphics{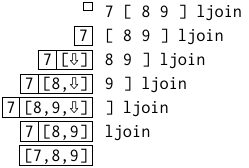}
\end{center}  
The arrow symbol shows where the next data item will be inserted.
This notation is more important when we have nested lists.
As we close the inner open list, the indicator arrow moves outwards.
\begin{center}
  \includegraphics{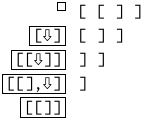}
\end{center}
We have the analogous operation \texttt{rjoin} and the inverses (on non-empty lists) \texttt{lcut} and \texttt{rcut}.

\subsection{Quoting and Dequoting}

The list is not just a collection, it is also used for \emph{quoting}.
Quoting works just like in natural language.
Compare the two sentences:

\begin{enumerate}
\item Cambridge is a nice university town in England.
\item``Cambridge'' starts with the letter C.
\end{enumerate}
In the first sentence, our attention turns to the town itself, imagining the medieval buildings, or punting on the river.
We move from the word to the meaning.
In computing terminology, we \emph{evaluate} the token.
In the second case, the evaluation is withheld.
We look at the token itself.
Similarly, putting code into a list makes it into inert data.
\begin{center}
  \includegraphics{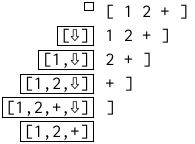}
\end{center}  
Nothing happens to the code, no arithmetic calculation is done - that is the point.
We pushed the code itself to the stack as data for delayed execution or manipulation.

How to quote something already on the stack? We simply need to apply previous list handling knowledge.
\begin{center}
\includegraphics{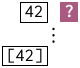}
\end{center}  
The program \texttt{[ ] ljoin} pushes an empty list on top, and by left-joining it swallows the previous top element.

The inverse operation of quoting, \emph{dequoting}, can be done by the \texttt{i} operator. 
It prepends the content of a program (a list) on the top of the stack to the input buffer.
Thus, the program gets executed.
Visually, it looks like that we move the code back to the input buffer in one move.
\begin{center}
  \includegraphics{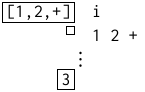}
\end{center}
The name \texttt{i} comes from combinatory logic \cite{hindley2008lambda, smullyan1985mock}, where $I$ stands for the identity combinator.
Perhaps the name is not most fitting, as dequoting renders inactive code active.

How to act deeper on the stack?
In the following puzzle we want to add \texttt{2} and \texttt{3}, but these operands are deeper down in the stack.
\begin{center}
  \includegraphics{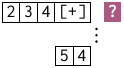}
\end{center}
We have the quoted program for addition and the number \texttt{4} is also in the way.
We can swap the code closer to the operands, then join the number with the addition program on the right.
Finally, we execute the prepared program.
Number \texttt{4} will push itself back to the top.
Therefore, the solution is creating a program that executes the given program and restores the stack: \texttt{swap rjoin i}.

\subsection{Homoiconicity and Gödel}

We saw that programs are represented as lists, sequences of tokens.
The input buffer itself is a list, although we omit the brackets in our notation.
Similarly  for the stack, we represent it graphically, but it is a list too.
This is an example of the \emph{code as data} principle.
In other words, \CC~is a \emph{homoiconic} language.

This property allows metaprogramming, which is the standard way of programming in \CC.
This is also a simplified version of Gödel-numbering \cite{GEB}.
We can easily express the idea of a program processing itself.
If \texttt{P} is any program that takes a list, then the following program is valid.
\begin{verbatim}
[ [ P ] P ]
\end{verbatim}

\subsection{Metalanguage}

The standard use of programming languages is to extend them with new function definitions.
For the original mathematical purpose of \CC, function definitions are redundant.
They do not increase the computational capabilities of the language, just make programs shorter.
However, for problem solving this feature is needed.
As a middle-way solution, we define a metalanguage layer by tokens that start with a dot, like hidden files in a UNIX filesystem.
\begin{center}
  \includegraphics{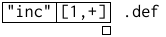}
\end{center}  
The effect is that a new token \texttt{inc} will be available for use.
The stack is empty and the real change in the language is invisible.
\begin{center}
  \includegraphics{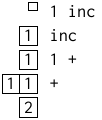}
\end{center}  
Processing a defined token is prepending the associated code to the input buffer.
This is the same mechanism as the one used for the \texttt{i} operator.
Note that the definition also requires to have \emph{string} literals.

\subsection{Conditionals}

Branching the execution based on a value is a necessary ingredient for universal computation.
It is a standard technique to use the logical values \texttt{true} and \texttt{false} to choose between alternatives. 
\begin{center}
  \includegraphics{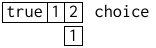}
\hskip55pt
  \includegraphics{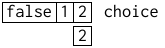}
\end{center}  
The order of condition, consequent and alternative can be arbitrary.
This convention follows the usual if-then-else format.
Based on \texttt{choice} and quoting we can define more sophisticated conditional statements.
For instance,
\begin{verbatim}
I T E ifte
\end{verbatim}
where \texttt{I}, \texttt{T}, \texttt{E} are programs.
First, the program \texttt{I} for condition is evaluated on the remaining stack.
It leaves a logical value on stack.
Based on that, we evaluate either \texttt{T} or \texttt{E}.
Other variations restore the stack after evaluating the condition.\footnote{See the standard library in \CC~source code, in the \texttt{resources} folder.}

\subsection{Recursion}
Once we can define new tokens, recursion is straightforward.
\begin{verbatim}
"factorial"
[ [ dup zero? ]
  [ drop 1 ]
  [ dup 1 - factorial * ]
  ifte ]
.def
\end{verbatim}
The \texttt{ifte} conditional first checks whether the top of the stack is zero or not.
The condition preventively duplicates the top number.
If true, we replace that with number one.
Otherwise, we push a decremented number on the stack.
This will also accumulate the required number of multiplication tokens.
No multiplication happens until the recursion stops.

Can we do recursion without names?
We can use the $Y$-combinator \cite{hindley2008lambda}.
It is even easier to express in a concatenative language \cite{joyrecursion}.
The idea is that in the moment of recursive call, we need to have access to source code of the recursive function.
In our case, any program is a list.
Therefore, it is only a matter of keeping a copy of that list on the stack until the base line case of recursion is called.
It does not need to be defined as a library token.

\subsection{Functional Programming}

Direct stack manipulation may make the impression that \CC~is only operating as low-level machine code.
However, with the expressive power of quoting it is easy to implement higher order functions used in most functional languages.
For instance, \texttt{map} is used for transforming a collection by a given function.
\begin{center}
  \includegraphics{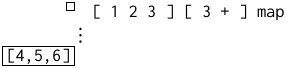}
\end{center}
Reducing a collection to a value is done by \texttt{fold}.
\begin{center}
  \includegraphics{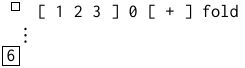}
\end{center}
We specify the collection to be reduced, the starting cumulative value, and a reducing function, here addition.

This situation is similar to the question whether the \textsc{Forth} language is low-level or high-level.
As argued in \cite{ThinkingForth}, the syntax is almost non-existent, resembling machine code, but the possibility for abstraction is not limited, so the language is flexible enough to serve those two purposes.

\section{Implementation and Trials}

\begin{figure*}
\begin{verbatim}
con-cat 2026.04.03 #02f154d Clojure 1.12.4 Java 25.0.3-ea+7-Ubuntu-2 Linux 7.0.0-1009-raspi aarch64
KERNEL   *  +  -  /  <  =  [  ]  and  cat  choice  count  drop  dup  empty?  get  i  id  last  lcut  ljoin  mod
newstack  not  or  rcut  rolldown  rollup  stack  stack-size  swap  unstack
LIBRARY   ->  any?  ddip  ddrop  dec  dip  drop4  duco  even?  factorial  factorial2  filter  first  fix  fold
ifte  inc  map  map2  odd?  over  pack  pair  quote  range  rec-fold  rec-map  rec-while  remove  rjoin  rjoin-if
run  rund  sifte  sim  square  sum  tdip  tdrop  times  while  y  zero?  |
META/SYSTEM   .def  .depth  .language  .load  .quit  .src

empty stack
1 2 +

0: 3
\end{verbatim}
\caption{The output of starting the \CC~interpreter. The complete set of tokens, the whole language, is printed. The tokens are categorized as kernel (fast and opaque), library (interpreted, traceable, source code available) and meta (system level operations with side-effects not visible on the stack). Note that this is a development snapshot, not the final specification of the language. An example arithmetic calculation is also shown. The top of stack is indexed by zero. The index grows with depth.}
\label{fig:concat}  
\end{figure*}

\subsection{The \CC-runtime}

\CC~is hosted on \textsc{Clojure} \cite{clojure2020}, which is a modern dialect of \textsc{Lisp}.
At this stage we focus on finding the right language primitives and correctness, and leave efficiency for a later stage.
This is a standard approach for experimental languages \cite{nystrom2021crafting}.
Using a high-level language makes the implementation straightforward: reusing data structures, avoiding memory management.
Since \CC~does not have variable, immutability is not directly a concern.

There is a distinction between \emph{kernel} and \emph{library} level tokens (Figure \ref{fig:concat}).
The formers are implemented in the host language, the latter are defined by \texttt{.def}.
Kernel code is faster, but opaque.
Moving between the two levels is possible by redefining kernel tokens.
The source code and documentation for the language is available at \cite{concat}.

There is evidence that the error messages in a programming language are crucial for the efficiency of learning \cite{2021errormessages}.
In an in-person event, it is the responsibility of the host to provide the right amount of feedback for the participants.
In addition, we also pay attention to the error messages in the implementation according to the guidelines \cite{2021errormessages}: avoiding jargon, and aiming for complete but economical sentences.

\subsection{Typesetting the Execution Traces}

All trace diagrams in this paper were generated by the \CC~ runtime environment \cite{concat}.
Using the same code that executes the programs guarantees the correctness of the examples.
The typographic layout was done by \textsc{Typst}, a modern, markup-based document preparation system \cite{2026typst}.
\textsc{Typst} has an easy to use scripting language, convenient for typesetting the execution traces. 
\subsection{Trials}
The content of the course material was tested by participants of diverse backgrounds.
They include schoolchildren, university students and professors, and software engineers.
Their feedback was instructive for designing the language and the typographic presentations.
The trials are still ongoing, therefore the language is not fixed yet.
\section{Conclusion}

Free from the constraints of practicality and compatibility, we could choose a programming style for simplicity and depth, aiming for the efficiency of learning.
Concatenative functional programming offers a unified view of machine code level processing and higher order functions, again shortcutting traditional layered computer science curriculum.
We developed a language, \CC, according to these guidelines.

\subsection{Future Work}

The project is still in the preparatory phase.
The next tasks will be extending the content and start measuring the efficiency.

We will adapt relevant content from the legendary SICP textbook \cite{SICP96}.
Another way to improve attractiveness is to implement \CC~on a retrocomputing device. The best-selling and recently reissued Commodore 64 microcomputer \cite{juul2024too} is a challenging but rewarding implementation target.

For measuring the efficiency of this teaching material, we will integrate the puzzles into extra-curricular mathematical activities.

\bibliographystyle{IEEEtran}
\bibliography{../coords.bib}

\end{document}